\documentclass[preprintnumbers, prd, floatfix, twocolumn, preprintnumbers, letterpaper, superscriptaddress,nofootinbib]{revtex4-1}

\usepackage{amsfonts}
\usepackage{amsmath}
\usepackage{amssymb,epsf}
\usepackage{latexsym}
\usepackage{graphicx,epsfig}
\usepackage{amssymb}
\usepackage{float}
\usepackage{subfigure}
\usepackage{epstopdf}
\usepackage[colorlinks=true,citecolor=blue,linkcolor=blue,urlcolor=black]{hyperref}
\usepackage{dcolumn}
\usepackage{psfrag}
\usepackage{wrapfig}
\usepackage{makeidx}
\usepackage{epsf}
\usepackage{color}
\usepackage{multirow}
\usepackage{mathtools}
\usepackage[normalem]{ulem}

\DeclarePairedDelimiter\abs{\lvert}{\rvert}
\newcommand{\be}{\begin{equation}}
\newcommand{\ee}{\end{equation}}
\newcommand{\bea}{\begin{eqnarray}}
\newcommand{\eea}{\end{eqnarray}}

\graphicspath{{Images/}}

\usepackage{tikz}

\definecolor{lime}{HTML}{A6CE39}
\newcommand{\orcidicon}{%
    \begin{tikzpicture}
    \draw[lime, fill=lime] (0,0)
        circle [radius=0.16]
        node[white] {{\fontfamily{qag}\selectfont \tiny ID}};
    \draw[white, fill=white] (-0.0625,0.095)
        circle [radius=0.007];
    \end{tikzpicture}   \hspace{-2mm}
}
\newcommand\orcidFrancisco{{\href{https://orcid.org/0000-0002-9388-8373}{\orcidicon}}}
\newcommand\orcidMahdi{{\href{https://orcid.org/0000-0003-1196-9493}{\orcidicon}}}

\begin{document}

\title{Dynamic wormhole geometries in hybrid metric-Palatini gravity}
\author{Mahdi Kord Zangeneh\orcidMahdi\!\!}
\email{mkzangeneh@scu.ac.ir}
\affiliation{Physics Department, Faculty of Science, Shahid Chamran University of Ahvaz,
Ahvaz 61357-43135, Iran}
\author{Francisco S. N. Lobo\orcidFrancisco\!\!}
\email{fslobo@fc.ul.pt}
\affiliation{Instituto de Astrof\'isica e Ci\^encias do Espa\c{c}o, Faculdade de
Ci\^encias da Universidade de Lisboa, Edif\'icio C8, Campo Grande,
P-1749-016, Lisbon, Portugal}

\date{\today}

\begin{abstract}
In this work, we analyse the evolution of time-dependent traversable wormhole geometries in a Friedmann-Lema\^{i}tre-Robertson-Walker background in the context of the scalar-tensor representation of hybrid metric-Palatini gravity. We deduce the energy-momentum profile of the matter threading the wormhole spacetime in terms of the background quantities, the scalar field, the scale factor and the shape function, and find specific wormhole solutions by considering a barotropic equation of state for the background matter. We find that particular cases satisfy the null and weak energy conditions for all times. In addition to the barotropic equation of state, we also explore a specific evolving wormhole spacetime, by imposing a traceless energy-momentum tensor for the matter threading the wormhole and find that this geometry also satisfies the null and weak energy conditions at all times.
\end{abstract}

\maketitle

\section{Introduction}

The explanation of the accelerated expansion of the Universe is one of the most challenging problems in modern cosmology \cite{Perlmutter:1998np,Riess:1998cb}. From the mathematical point of view, the simplest way to treat this problem is to consider the cosmological constant term  \cite{Carroll:2000fy}. Nonetheless, this model faces some difficulties such as the coincidence problem and the cosmological constant problem.
The latter dictates a huge discrepancy between the observed values of the vacuum energy density and the theoretical large value of the zero-point energy suggested by quantum field theory \cite{Carroll:2000fy}.
There are some alternative models proposed to overcome these problems such as modified gravity \cite{Nojiri:2010wj,Sotiriou:2008rp,Capozziello:2011et,Avelino:2016lpj,Lobo:2008sg,Bamba:2015uma,Nojiri:2017ncd}, mysterious energy-momentum sources \cite{Copeland:2006wr,Horndeski:1974wa,Deffayet:2011gz}, such as quintessence 
\cite{Wetterich:1987fm,Ratra:1987rm,Caldwell:1997ii,Zlatev:1998tr} and $k$-essence \cite{ArmendarizPicon:1999rj,ArmendarizPicon:2000dh,ArmendarizPicon:2000ah} fields, and complex equations of state storing the missing energy of the dark side of the Universe \cite{Bento:2002ps,Arun:2017uaw}.
In general, models with varying dark energy candidates may be capable of overcoming all the mathematical and theoretical difficulties, however, the main underlying question is the origin of these terms.
One proposal is to relate this behavior to the energy of quantum fields in vacuum through the holographic principle which allows us to reconcile infrared (IR) and ultraviolet (UV) cutoffs \cite{Cohen:1998zx} (see Ref.~\cite{Tawfik:2019dda} for a review on various attempts to model the dark side of the Universe).

However, an alternative to these dark energy models, as mentioned above, is modified gravity \cite{Nojiri:2010wj,Sotiriou:2008rp,Capozziello:2011et,Avelino:2016lpj,Lobo:2008sg,Bamba:2015uma,Nojiri:2017ncd}. Here, one considers generalizations of the Hilbert-Einstein Lagrangian specific curvature
invariants such as $R^{2}$, $R_{\mu \nu }R^{\mu \nu }$, $R_{\alpha
\beta \mu \nu }R^{\alpha \beta \mu \nu }$, $\varepsilon ^{\alpha
\beta \mu \nu }R_{\alpha \beta \gamma \delta }R_{\mu \nu }^{\gamma
\delta }$, etc. A popular theory that has attracted much attention is $f(R)$ gravity, where one may tackle the problem through several approaches, namely, the metric formalism \cite{Nojiri:2010wj,Sotiriou:2008rp,Capozziello:2011et,Avelino:2016lpj,Lobo:2008sg,Bamba:2015uma,Nojiri:2017ncd}, which considers that the metric is the fundamental field, or the Palatini formalism \cite{Olmo:2011uz}, where here one varies the action with respect to the metric and an independent connection. However, one may also consider a hybrid combination of these approaches that has recently been proposed, namely, the hybrid metric-Palatini gravitational theory \cite{Harko:2011nh}, where the metric Einstein-Hilbert action is supplemented with a metric-affine (Palatini) correction term. The hybrid metric-Palatini theory has the ability to avoid several of the problematic issues that arise in the pure metric and Palatini formalisms. 
For instance, metric $f(R)$ gravity introduces an additional scalar degree of freedom, which must possess a low mass in order to be relevant for the large scale cosmic dynamics. However, the presence of such a low mass scalar field would influence the dynamics on smaller scales, such as at the level of the Solar System, and since these small scale effects remain undetected, one must resort to screening mechanisms \cite{Capozziello:2007eu,Khoury:2003rn}. Relative to the Palatini formalism, no additional degrees of freedom are introduced, as the scalar field is an algebraic function of the trace of the energy-momentum tensor. It has been shown that this fact entails serious consequences for the theory, leading to the presence of infinite tidal forces on the surface of massive astrophysical type objects \cite{Olmo:2011uz}.

In this context, the hybrid metric-Palatini gravity theories were proposed initially in \cite{Harko:2011nh} in order to circumvent the above shorcomings in the metric and Palatini formalisms of $f(R)$ gravity. One of the main advantages of the hybrid metric-Palatini theory is that in its scalar-tensor representation a long-range force is introduced that automatically passes the Solar System tests, and thus no contradiction between the theory and the local measurements arise. In a cosmological context, it has also been shown that hybrid metric-Palatini gravity may also explain the cosmological epochs \cite{Lima:2014aza,Lima:2015nma}
(we refer the reader to \cite{Capozziello:2015lza,Harko:2018ayt,Harko:2020ibn} for more details).
In fact, hybrid metric-Palatini gravity has attracted much attention recently, and has a plethora of applications, namely, in the galactic context \cite{Capozziello:2012qt,Capozziello:2013yha,Capozziello:2013uya}, in cosmology \cite{Capozziello:2012ny,Carloni:2015bua,Capozziello:2013wq,Boehmer:2013oxa,Lima:2014aza,Lima:2015nma}, in braneworlds \cite{Fu:2016szo,Rosa:2020uli}, black holes \cite{Bronnikov:2019ugl,Chen:2020evr,Rosa:2020uoi,Bronnikov:2020vgg}, stellar solutions \cite{Danila:2016lqx}, cosmic strings \cite{Bronnikov:2020zob,Harko:2020oxq}, tests of binary pulsars \cite{Avdeev:2020jqo}, and wormholes \cite{Rosa:2018jwp,Capozziello:2012hr}, amongst others.

Here, we explore the possibility that evolving wormhole geometries may be supported by hybrid metric-Palatini gravity. These compact objects, which are theoretical shortcuts in spacetime, have been shown to be threaded by an exotic fluid that violates the null energy condition (NEC), at least for the static case \cite{Morris:1988cz,Morris:1988tu}. However, it has been shown that evolving wormholes are able to satisfy the
energy conditions in arbitrary but finite intervals of time \cite{Kar:1994tz,Kar:1995ss}, contrary to their static counterparts \cite{Visser:1995cc,Lobo:2017oab}. One way to study this subject is to embed a wormhole in a Friedmann–Lema\^{i}tre–Robertson–Walker (FLRW) metric, which
permits the geometry to evolve in a cosmological background \cite{Roman:1992xj,Anchordoqui:1997du,Arellano:2006ex,Ebrahimi:2009ih,Ebrahimi:2009rv,Bordbar:2010zz,Sajadi:2012zz,Cataldo:2013sma,Setare:2015iqa, Bhattacharya:2015oma,Ovgun:2018uin,Zangeneh:2014noa,Cataldo:2008pm,Cataldo:2008ku,Cataldo:2012pw,Maeda:2009tk,KordZangeneh:2020jio}. Due to the somewhat problematic nature of the energy condition violations, an important issue is whether traversable wormholes can be constructed from normal matter throughout the spacetime, or at least partially. Different wormhole structures have been explored from this point of view \cite{Zangeneh:2014noa,Mehdizadeh:2015jra,Mehdizadeh:2015dta,Zangeneh:2015jda,KordZangeneh:2020jio,Parsaei:2019utg}. Indeed, it has been shown that in modified gravity, it is possible to impose that the matter threading the wormhole throat satisfies the energy conditions \cite{Capozziello:2013vna,Capozziello:2014bqa}, and it is the higher order curvature terms that support these nonstandard wormhole geometries \cite{Lobo:2007qi,Lobo:2009ip,Garcia:2010xb,MontelongoGarcia:2010xd,Harko:2013yb,Korolev:2020ohi}. Furthermore, different wormhole structures have been studied extensively in the context of alternative theories of gravity \cite{Antoniou:2019awm, Tangphati:2019pxh, Papantonopoulos:2019ugr, Godani:2019kgy, Banerjee:2020uyi, Fayyaz:2020jzh, Restuccia:2020wls,Tangphati:2020mir,Singh:2020rai,Ibadov:2020btp,Richarte:2009zz} and also from different aspects \cite{Lazov:2017tjs, Savelova:2019lye, Bak:2019nnu, Xu:2020wfm, Jusufi:2020rpw, Berry:2020tky, Fallows:2020ugr, Maldacena:2020sxe,Lobo:2020kxn,Parsaei:2020hke,DeFalco:2020afv,Moti:2020whf,Wielgus:2020uqz}.

In this paper, we study the evolution of traversable wormholes in a FLRW
universe background in the scalar-tensor representation of hybrid metric-Palatini gravity.
Furthermore, we explore the energy conditions for matter which threads these
wormhole geometries. The paper is organized in the following manner: In Sec. \ref{secII}, we briefly present the action and field equations of hybrid metric-Palatini gravity, we consider the spacetime metric and explore a barotropic equation of state for the background fluid. In Sec. \ref{secIII}, we analyse evolving traversable wormhole geometries for specific values of the barotropic equation of state parameter, as well as evolving wormholes with a traceless energy-momentum tensor (EMT), and study the energy conditions for the solutions obtained. Finally, in Sec. \ref{conclusion}, we summary our results and conclude.

\section{Evolving wormholes in hybrid metric-Palatini gravity} \label{secII} 

\subsection{Action and field equations}

Here, we briefly present the hybrid metric-Palatini gravitational theory. The action is given by
\begin{equation}  \label{eq:S_hybrid}
S= \frac{1}{2\kappa^2}\int d^4 x \sqrt{-g} \left[ R + f(\mathcal{R})\right]
+S_m \ ,
\end{equation}
where $\kappa^2\equiv 8\pi G$, $R$ is the metric Ricci scalar, and the Palatini curvature is $\mathcal{R} \equiv g^{\mu\nu}\mathcal{R}_{\mu\nu} $, with the Palatini Ricci tensor, $\mathcal{R}_{\mu\nu}$, defined in terms of an independent connection, $\hat{\Gamma}^\alpha_{\mu\nu}$, given by
\begin{equation}
\mathcal{R}_{\mu\nu} \equiv \hat{\Gamma}^\alpha_{\mu\nu ,\alpha} - \hat{%
\Gamma}^\alpha_{\mu\alpha , \nu} + \hat{\Gamma}^\alpha_{\alpha\lambda}\hat{%
\Gamma}^\lambda_{\mu\nu} -\hat{\Gamma}^\alpha_{\mu\lambda}\hat{\Gamma}%
^\lambda_{\alpha\nu}\,,
\end{equation}
and $S_m$ is the matter action. 

However, the scalar-tensor representation of hybrid metric-Palatini gravity provides a theoretical framework which is easier to handle from a computational point of view \cite{Harko:2011nh}, where the equivalent action   to (\ref{eq:S_hybrid}) is provided by 
\begin{equation}  \label{eq:S_scalar2}
S=\int \frac{d^4 x \sqrt{-g} }{2\kappa^2}\left[ (1+\phi)R +\frac{3}{2\phi}%
\partial_\mu \phi \partial^\mu \phi -V(\phi)\right]+S_m .
\end{equation}
Note a similarity with the the action of the $w=-3/2$ Brans-Dicke theory version of the Palatini approach to $f(R)$ gravity. However, the hybrid theory exhibits an important and subtle difference appearing in the scalar field-curvature coupling, which in the $w=-3/2$ Brans-Dicke theory is of the form $\phi R$ \cite{Harko:2011nh}.

By varying the action (\ref{eq:S_scalar2}) with respect to the metric provides the following gravitational field equation
\begin{equation}
G_{\mu \nu }=\kappa ^{2}\left( \frac{1}{1+\phi }T_{\mu \nu }+T_{\mu \nu
}^{(\phi )}\right) ,  \label{einstein_phi}
\end{equation}%
where $T_{\mu \nu }$ is the standard matter energy-momentum tensor, and
\begin{eqnarray}
T_{\mu \nu }^{(\phi )} &=&\frac{1}{\kappa ^{2}}\frac{1}{1+\phi }\Bigg[\nabla
_{\mu }\nabla _{\nu }\phi -\frac{3}{2\phi }\nabla _{\mu }\phi \nabla _{\nu
}\phi  \notag \\
&&\text{\hskip0.75cm}+\Bigg(\frac{3}{4\phi }\nabla _{\lambda }\phi \nabla
^{\lambda }\phi -\square \phi -\frac{1}{2}V\Bigg)g_{\mu \nu }\Bigg] ,
\label{tens_perfect}
\end{eqnarray}%
is the energy-momentum tensor of the scalar field of the theory.

Varying the action with the scalar field yields the following second-order differential equation, 
\begin{equation}  \label{eq:evol-phi}
-\square \phi+\frac{1}{2\phi}\partial_\mu \phi \partial^\mu \phi +\frac{\phi[%
2V-(1+\phi)V_\phi]} {3}=\frac{\phi\kappa^2}{3}T \ .
\end{equation}
The above equation shows that in hybrid metric-Palatini gravity the scalar field is dynamical. This
represents an important and interesting difference with respect to the
standard Palatini case \cite{Olmo:2011uz}.

\subsection{Spacetime metric}

The metric of a time-dependent wormhole geometry is given by
\begin{equation}
ds^{2}=-e^{2\Phi \left( r\right) }dt^{2}+a^2\left( t\right)\left[ \frac{%
dr^{2}}{1-b\left( r\right) /r}+r^{2}d\Omega ^{2}\right] ,  \label{met}
\end{equation}%
where $\Phi \left( r\right)$ and $b\left(r\right) $, respectively, are the redshift and shape functions, which are $r$-dependent, and $a\left( t\right) $ is the time-dependent scale factor, and $d\Omega ^{2}=d\theta ^{2}+\sin^{2} \theta d\varphi ^{2}$ is the linear element of the unit sphere. 

To correspond to a wormhole solution, the shape function has the following restrictions \cite{Morris:1988cz}: (i) $b(r_{0})=r_{0}$, where $r_{0}$ is the wormhole throat, corresponding to a minimum radial coordinate, (ii) $b(r)\leq r$, and (iii) the flaring-out condition given in the form $rb^{\prime }(r)-b\left( r\right) <0$. The latter condition is the fundamental ingredient in wormhole physics, as taking into account the Einstein field equation, one verifies that this flaring-out condition imposes the violation of the NEC. In fact, it violates all of the
pointwise energy conditions \cite{Morris:1988cz,Morris:1988tu,Visser:1995cc,Lobo:2017oab,Lobo:2004wq}. In order to avoid the presence of event horizons, so that the wormhole is traversable, one also imposes that the redshift function, $\Phi(r)$, be finite everywhere. However, throughout this work, we consider a zero redshift function, $\Phi =0$, which simplifies the calculations significantly.

\subsection{Evolving embedding analysis}

It is also interesting to explore how the time evolution of the scale factor affects the wormhole itself. To this effect, in the following analysis, we will follow closely the analysis outlined in Refs. \cite{Morris:1988cz,Roman:1992xj,KordZangeneh:2020jio}, and analyse how the (effective) radius of the throat or length of the wormhole changes it time. As mentioned above, throughout this paper, we consider $\Phi =0$ for simplicity. Thus, in order to analyse the time-dependent dynamic wormhole geometry, we need to choose a specific $b(r)$ to provide a reasonable wormhole at $t=0$, which is assumed to be the onset of the evolution. Note that the radial proper length between any two points $A$ and $B$, through the wormhole, at any $t={\rm const}$
is given by $l(t)=\pm \, a(t) \int_{r_A}^{r_B} {(1-b/r)^{-1/2}}\,dr$, which is merely the initial radial proper separation multiplied by the scale factor.

Now, to analyse how the ``wormhole'' form of the metric is maintained throughout the evolution, we consider a $t={\rm const}$ and $\theta=\pi/2$ slice of the spacetime (\ref{met}). Thus, the metric reduces to
\begin{equation}\label{slice}
ds^2=\frac{a^2(t)\,dr^2}{1-b(r)/r} + a^2(t)\,r^2\,d\varphi^2\,,
\end{equation}
and is embedded in a flat 3-dimensional Euclidean space given by
\begin{equation}
ds^2=d{\bar{z}}^2+d{\bar{r}}^2+{\bar{r}}^2\,{d\varphi}^2\,.
\label{barredslice}
\end{equation}
Comparing the angular coefficients, yields
\begin{eqnarray}
\bar{r}&=&{a(t)\,r}\big|_{t={\rm const}} \,,
       \label{coef1:phi}      \\
{d\bar{r}}^2&=&a^2(t)\,{dr}^2\big|_{t={\rm const}}  \,.
\label{coef2:phi}
\end{eqnarray}
Note, that when considering derivatives, these relations do not represent a ``coordinate transformation'', but a ``rescaling'' of the radial coordinate $r$ along each $t={\rm constant}$ slice.

The ``wormhole'' form of the metric will be preserved, with respect to the ${\bar{z}},{\bar{r}},\varphi$ coordinates, if the embedded slice has the following metric
\begin{equation}\label{WHslice}
ds^2=\frac{d{\bar{r}}^2}{1-\bar{b}(\bar{r})/\bar{r}} +
                      {\bar{r}}^2{d\varphi}^2\,,
\end{equation}
and the shape functions $\bar{b}(\bar{r})$ has a minimum at a specific $\bar{b}(\bar{r}_0)=\bar{r}_0$. Now, the embedded slice  (\ref{slice}) can be readily rewritten in the form of Eq. (\ref{WHslice}) by using Eqs. (\ref{coef1:phi}) and (\ref{coef2:phi}), and the following relation
\begin{equation}
\bar{b}(\bar{r})=a(t)\,b(r).
    \label{bar:b}
\end{equation}
Thus, the evolving wormhole will have the same overall shape and size and relative to the ${\bar{z}},{\bar{r}},\varphi$ coordinate system, as the initial wormhole had relative to the initial $z,r,\varphi$ embedding space coordinate system. 
In addition to this, using Eqs. (\ref{barredslice}) and (\ref{WHslice}), one deduces that
\begin{equation}
{{d{\bar{z}}}\over{d{\bar{r}}}}
=\pm\left({{\bar{r}}\over{\bar{b}(\bar{r})}}-1\right)^{-1/2}
={{dz}\over{dr}} \,,
            \label{barredembedding}
\end{equation}
which implies
\begin{eqnarray}
\bar{z}(\bar{r})&=&\pm\int{{d\bar{r}}\over{(\bar{r}/{\bar{b}(\bar{r})}-1)^{1/2}}}
	\nonumber \\
&=& \pm \, a(t)\,\int{\left(\frac{r-b}{b}\right)^{-1/2}} \,dr
	\nonumber  \\
&=&\pm \, a(t)\,z(r)\,.
            \label{embed:relation}
\end{eqnarray}
Thus, the relation between the initial embedding space at $t=0$ and the embedding space at any time $t$ is, from Eqs. (\ref{coef2:phi}) and (\ref{embed:relation})
\begin{equation}
ds^2=d{\bar{z}}^2+d{\bar{r}}^2+{\bar{r}}^2\,{d\varphi}^2
        =a^2(t)\,[dz^2+dr^2+r^2{d\varphi}^2]\,.
\end{equation}
The wormhole will always remain the same size, relative to the ${\bar{z}},{\bar{r}},\phi$ coordinate system, as the scaling of the embedding space compensates for the evolution of the wormhole.
Nevertheless, the wormhole will change size relative to the initial $t=0$ embedding space.

We also note that the ``flaring out condition'' for the evolving wormhole is given by
\begin{equation}
{{d\,^2{\bar{r}(\bar{z})}}\over{d{\bar{z}}^2}}>0\,,
\end{equation}
at or near the throat, so that taking into account Eqs. (\ref{coef1:phi}),
(\ref{coef2:phi}), (\ref{bar:b}), and (\ref{barredembedding}), it
follows that
\begin{equation}
{{d\,^2{\bar{r}(\bar{z})}}\over{d{\bar{z}}^2}}
       =\frac{1}{a(t)}\,{{b-b'r}\over{2b^2}}
       =\frac{1}{a(t)}\,{{d\,^2r(z)}\over{dz^2}}>0\,,
       \label{barred:flareout}
\end{equation}
at or near the throat. Using Eqs. (\ref{coef1:phi}),
(\ref{bar:b}), and
\begin{equation}
{\bar{b}}'(\bar{r})={{d\bar{b}}\over{d\bar{r}}}
          =b'(r)={{db}\over{dr}}\,,
\end{equation}
the right-hand side of Eq. (\ref{barred:flareout}), relative to the ${\bar{z}},{\bar{r}},\phi$ coordinate system, may be written as
\begin{equation}
{{d\,^2{\bar{r}(\bar{z})}}\over{d{\bar{z}}^2}}
=\left({{\bar{b}-{\bar{b}}'\bar{r}}\over{2{\bar{b}}^2}}\right)>0\,,
         \label{barred:flareout2}
\end{equation}
at or near the throat. Thus, we have verified that the flaring out condition (\ref{barred:flareout2}), using the barred coordinates, has the same form as for the static wormhole.

\subsection{Gravitational field equations}

We also consider an anisotropic matter energy-momentum tensor given by $T_{\nu }^{\mu }={\rm diag}(-\rho,-\tau, p, p)$, where $\rho $, $\tau $
and $p$ are the energy density, the radial tension (which is equivalent to a negative radial pressure) and the tangential pressure, respectively. Using the metric (\ref{met}), the gravitational
field equations (\ref{einstein_phi}) provide the following energy-momentum profile:
\begin{equation}
\rho (t,r) =\left( 3H^{2} +\frac{b^{\prime }}{r^{2}a^{2}}%
\right) \left( 1+\phi \right) +3\dot{\phi}H+\frac{3\dot{\phi}^{2}}{4\phi }-%
\frac{1}{2}V(\phi ), 
~  \label{rhoeq} 
\end{equation}
\begin{eqnarray}
\tau \left( t,r\right) &=&\left( H^{2}+2\frac{\ddot{a}}{a} +\frac{b}{%
r^{3}a^{2}}\right) \left( 1+\phi \right)  \notag \\
&&+2\dot{\phi}H-\frac{3\dot{\phi}^{2}}{4\phi }+\ddot{\phi}%
-\frac{1}{2}V(\phi ),
\end{eqnarray}
\begin{eqnarray}
p\left( t,r\right) &=&\left( -H^{2}-2\frac{\ddot{a}}{a} -\frac{rb^{\prime}-b}{%
2r^{3}a^{2}}\right) \left( 1+\phi \right)  \notag \\
&&-2\dot{\phi}H+\frac{3\dot{\phi}^{2}}{4\phi }-\ddot{\phi}%
+\frac{1}{2}V(\phi ),  \label{pteq}
\end{eqnarray}
respectively, where $H=\dot{a}/a$ and $\phi =\phi \left( t\right) $, the overdot and prime denote derivatives with respect to $t$ and $r$, respectively, and for notational simplicity, we have considered $\kappa =1$. One recovers the standard field equations of the Morris-Thorne wormhole \cite{Morris:1988cz}, by fixing the scale factor $a$ to unity and excluding the background time-dependent evolution and the scalar field $\phi $ contribution.

Furthermore, we will also analyse the null and weak energy conditions for the solutions obtained below. The weak energy condition (WEC) is defined as $T_{\mu \nu
} u^{\mu }u^{\nu }\geq 0$, where $u^{\mu }$ is a timelike vector, and is
expressed in terms of the energy density $\rho $, radial tension $\tau $ and
tangential pressure $p$ as $\rho \geq 0$, $\rho -\tau \geq 0$ and $\rho
+p\geq 0$, respectively. The last two inequalities, i.e., $\rho -\tau \geq 0$
and $\rho +p\geq 0$ correspond to the NEC, which is defined as $T_{\mu \nu
}k^{\mu }k^{\nu }\geq 0$, where $k^{\mu }$ is \textit{any} null vector.

The scalar field equation of motion (\ref{eq:evol-phi}) yields
\begin{equation}
\ddot{\phi}+3\dot{\phi}H-\frac{\dot{\phi}^{2}}{2\phi }-\frac{1}{3}\phi \left[
T +\left( 1+\phi \right) \frac{dV}{d\phi }-2V(\phi )\right]
=0,  \label{phieom}
\end{equation}%
where the trace of the energy-momentum tensor $T=T\left( t,r\right) =T_{\mu }^{\mu }=-\rho -\tau +2p$ is given by
\begin{eqnarray}
T\left( t,r\right) = -2\left( \frac{b^{\prime }}{a^{2}r^{2}}+3H^{2}+3\frac{%
\ddot{a}}{a}\right) (1+\phi )  \notag \\
-3\ddot{\phi}+\frac{3\dot{\phi}^{2}}{2\phi }-9\dot{\phi}%
H+2V(\phi ).  \label{trace}
\end{eqnarray}%
In order to solve the differential equation (\ref{phieom}) for $\phi \left(
t\right) $, the trace $T\left( t,r\right) $, given by Eq. (\ref{trace}),
should be independent of $r$.  This condition leads to $b^{\prime }/r^{2}=CH_{0}^{2}$
where $H_{0}$ is the present value of the Hubble parameter and{\ $C$ is an
arbitrary dimensionless constant. Thus, the shape function is given by $b(r)=r_{0}+C{H_{0}^{2}}%
(r^{3}-r_{0}^{3})/3$, which satisfies $b(r_{0})=r_{0}$, and the flaring-out condition at the throat imposes $C H_{0}^{2} r_{0}^{2} < 1$.

To solve Eq. (\ref{phieom}) numerically for $\phi$, it is useful to rewrite it in terms of dimensionless functions of the scale factor $a$. Thus, we consider the definitions: 
\begin{eqnarray}
\ddot{\phi}=\ddot{a}\phi ^{\prime }(a)+\dot{a}^{2}\phi ^{\prime \prime }(a), \qquad  \dot{\phi}=\dot{a}\phi ^{\prime }(a),  
	\nonumber \\
\ddot{a}=H\dot{a}+\dot{H}a, \qquad  \dot{H}=\dot{a}H^{\prime }(a), \qquad \dot{a}=Ha	 \,.
\nonumber
\end{eqnarray}
Note that in the above definitions, the prime denotes a derivative with respect to the scale factor. In addition to this, consider $U=V/3H_{0}^{2}$
and $E=H/H_{0}$, so that Eq. (\ref{phieom}) finally takes the following form
\begin{eqnarray}
\phi ^{\prime \prime }(a)-\frac{\phi ^{\prime 2}}{2\phi (a)}+\frac{%
4(a\phi ^{\prime }(a) + \phi(a))}{a^2}+\frac{E^{\prime }(a)\phi ^{\prime }(a)}{E(a)}  \notag
 \\
+\frac{2C\phi (a)}{3a^{4}E(a)^{2}}+\frac{2\phi
(a)E^{\prime }(a)}{aE(a)}-\frac{\phi (a)}{a^{2}E(a)^{2}}\frac{dU}{d\phi } =0. 
\label{phidleom}
\end{eqnarray}

\subsection{Barotropic equation of state}

To solve the differential equation (\ref{phidleom}) for $\phi \left( a\right) $, we need to deduce $E$. To this effect, it is also useful to define the following background quantities:
\begin{equation}
\rho _{b}\left( t\right) =3H^{2}, \qquad  \tau _{b}\left( t\right) =H^{2}+2\frac{\ddot{a}}{a}\,,
\end{equation}
and consider a background barotropic equation of state given by $\tau _{b}=-\omega _{b}\rho_{b} $. From this condition, we find $E=a^{-3\left( \omega _{b}+1\right) /2}$. One could solve this equation of state for scale factor as a function of the time coordinate $t$, which yields $a(t) \propto t^{2/3(1+\omega_{b})}$. For $\omega_{b} = -1$, the scale factor takes the exponential form in terms of $t$.
Furthermore, we set $C=0$ which satisfies the flaring-out condition at the throat (see discussion below Eq. (\ref{trace})), and in addition simplifies the solution, so that $b(r)=r_{0}$. 
Thus, taking into these conditions and using the dimensionless definitions given above, Eqs. (\ref{rhoeq})-(\ref{pteq}) lead to the following relations 
%
\begin{equation}
\frac{\rho }{3H_{0}^{2}} =
\frac{ 1+\phi (a)}{a^{3\left( \omega _{b}+1\right) }} 
+\frac{\phi ^{\prime }(a)%
}{a^{3\omega _{b}+2}}
+\frac{\phi ^{\prime 2}(a)}{4a^{3\omega _{b}+1}\phi (a)}-\frac{1}{2}U(\phi
), \label{rho}
\end{equation}
\begin{eqnarray}
&&\frac{\rho -\tau }{3H_{0}^{2}} =
\left( \frac{\omega _{b}+1}{a^{3\left(\omega _{b}+1\right) }}-\frac{r_{0}}{3a^{2}H_{0}^{2}r^{3}}\right) \left( 1+\phi(a)\right)
  \nonumber \\
&&\qquad +\frac{\left(\omega _{b}+1 \right) \phi ^{\prime }(a)}{2a^{3\omega _{b}+2}}%
+\frac{\phi ^{\prime 2}(a)}{2a^{3\omega _{b}+1}\phi (a)}-\frac{\phi ^{\prime
\prime }(a)}{3a^{3\omega _{b}+1}},  \label{rhotau}
\end{eqnarray}
\begin{eqnarray}
&&\frac{\rho +p}{3H_{0}^{2}} =
\left( \frac{\omega _{b}+1}{a^{3\left( \omega_{b}+1\right) }}+\frac{r_{0}}{6a^{2}H_{0}^{2}r^{3}}\right) \left( 1+\phi (a)\right)
\nonumber \\ && 
\qquad +\frac{\left(\omega _{b}+1 \right) \phi ^{\prime }(a)}{2a^{3\omega _{b}+2}}%
+\frac{\phi ^{\prime 2}(a)}{2a^{3\omega _{b}+1}\phi (a)}-\frac{\phi ^{\prime
\prime }(a)}{3a^{3\omega _{b}+1}},  \label{rhop}
\end{eqnarray}%
respectively. It is interesting to note that $\rho$ is independent of the radial
coordinate $r$. To keep the terms dimensionless, we will consider the wormhole throat as $r_{0}= A H_{0}^{-1}$, where $A$ is a dimensionless constant. In what follows, we set $A$ to unity.

In the following, we shall consider specific solutions for the parameters $\omega _{b}=(-1, 1/3,0)$, respectively. These parameters range tentatively correspond to the inflationary, radiation and matter epochs.

\section{Specific evolving wormhole solutions and the energy conditions}\label{secIII}

\begin{figure*}[htb!]
\centering%
\subfigure[~Specific case of $\omega_b=-1$  with $U(\phi)=\phi^{0.5}$] {\label{fig1a}%
\includegraphics[width=.32\textwidth]{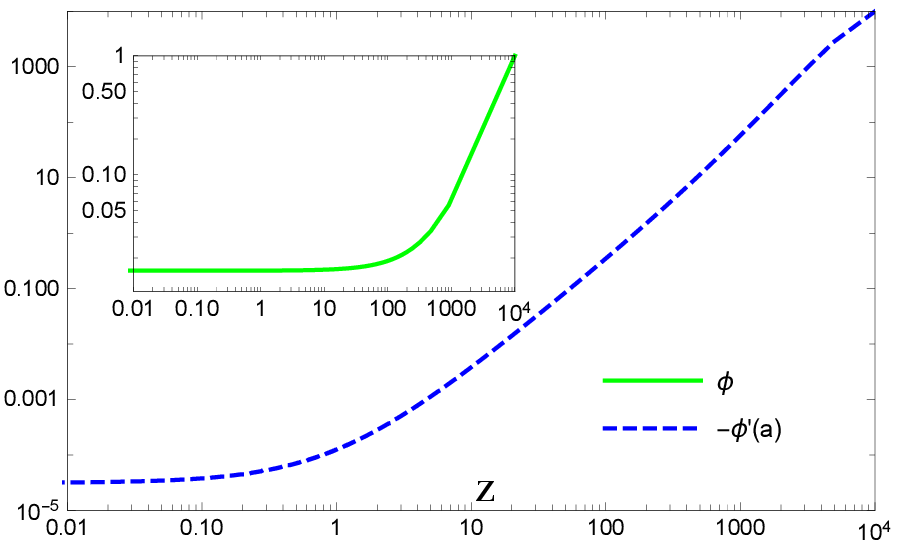} }
\subfigure[~Specific case of $\omega_b=1/3$ with
$U(\phi)=\phi^{2}$] {\label{fig1b}\includegraphics[width=.32\textwidth]{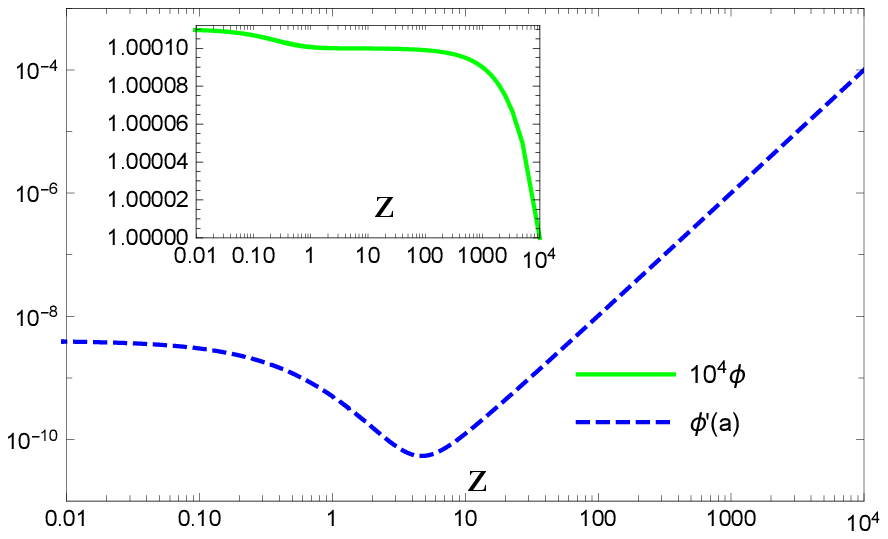} }
\subfigure[~Specific case of $\omega_b=0$ with $U(\phi)=\phi^{2}$] {\label{fig1c}\includegraphics[width=.32\textwidth]{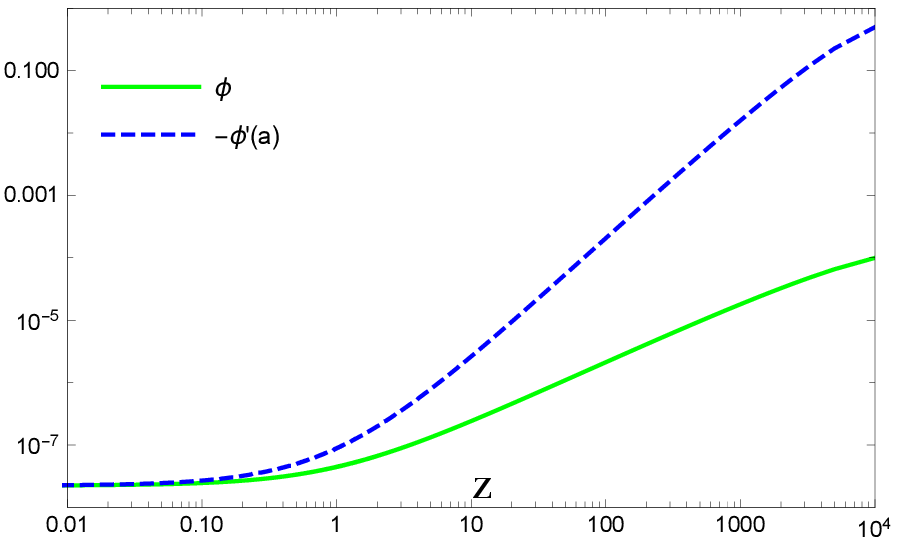}}
\caption{The behaviors of $\phi$ and $\phi'(a)$ vs $z$ for the specific case of  $\omega_b=-1$, $\omega_b=1/3$, and $\omega_b=0$, respectively. Note that both horizontal and vertical axes are logarithmic.}
\label{fig1}
\end{figure*}

\begin{figure*}[htb!]
\centering\subfigure[~$\rho/3H_0^2$ vs $z$] {\label{fig2a}%
\includegraphics[width=.32\textwidth]{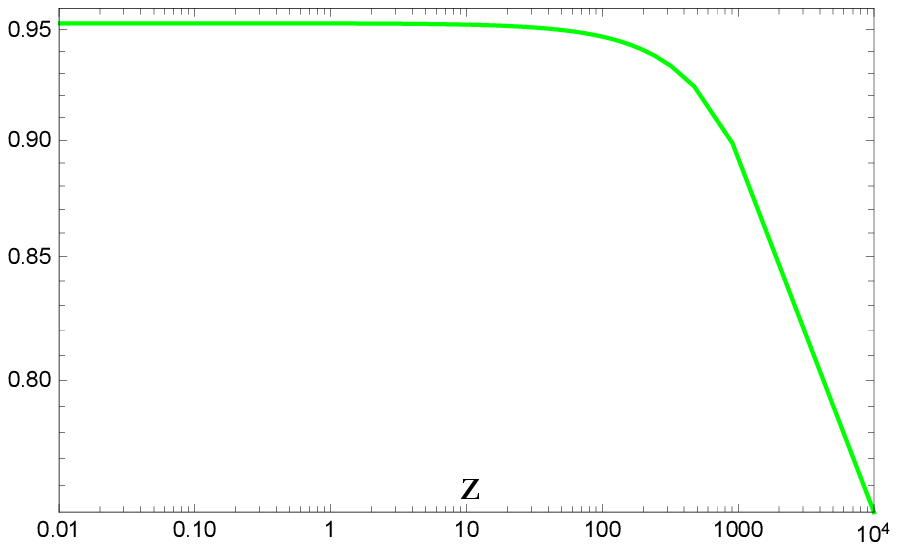} } 
\subfigure[~$ \abs{\rho-\tau}/3H_0^2$ vs $z$] {\label{fig2b}\includegraphics[width=.32\textwidth]{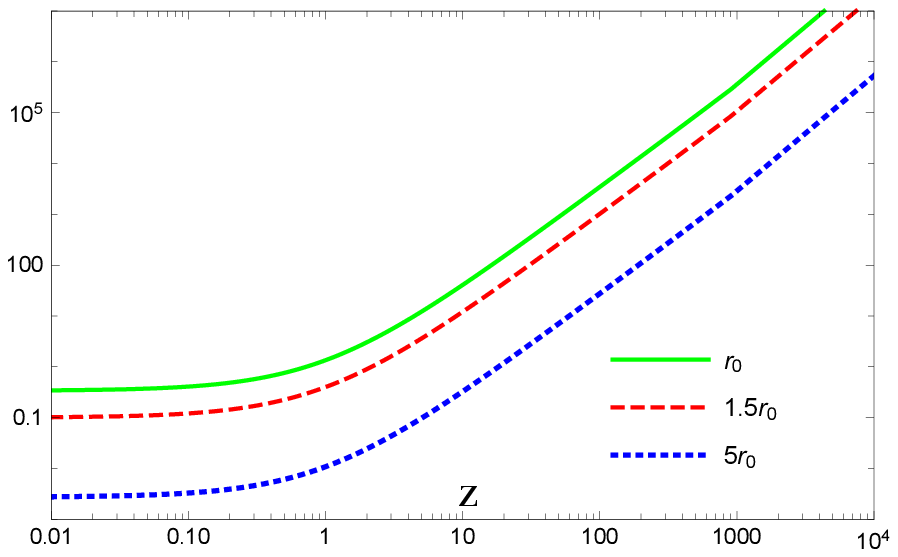}}
\subfigure[~$(\rho+p)/3H_0^2$ vs $z$] {\label{fig2c}%
\includegraphics[width=.32\textwidth]{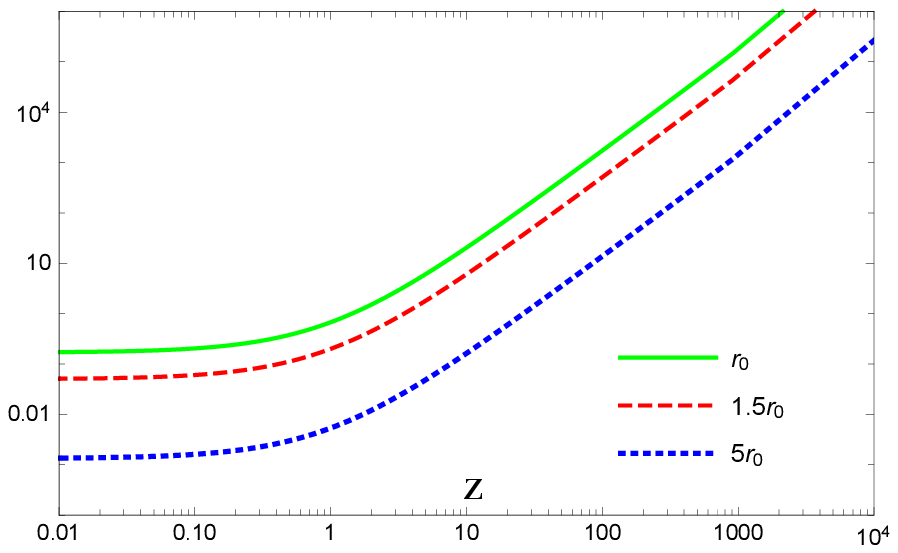} }
\caption{The behaviors of $\protect\rho $, $\protect\rho -\protect\tau $ (which is negative) and 
$\protect\rho +p$, respectively, versus $z$ for different values of $r$ for $\protect\omega_b=-1$ with $U(\protect\phi)=\protect\phi^{0.5}$. Note
that both horizontal and vertical axes are logarithmic, except for the vertical axis of Fig. \ref{fig2a}.}
\label{fig2}
\end{figure*}

\begin{figure*}[htb!]
\centering
\subfigure[~$\rho/3H_0^2$ vs $z$] {\label{fig3a} \includegraphics[width=.32\textwidth]{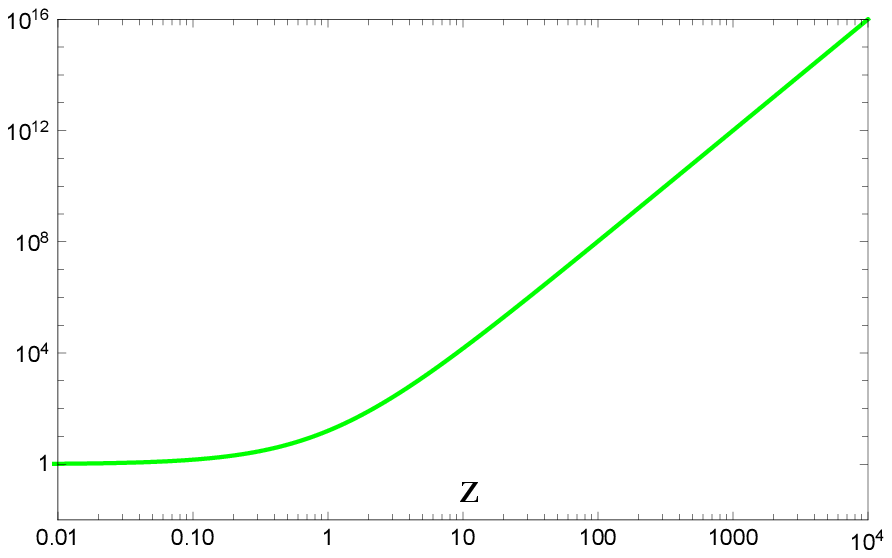}} 
\subfigure[~$(\rho-\tau)/3H_0^2$ vs $z$] {\label{fig3b}\includegraphics[width=.32\textwidth]{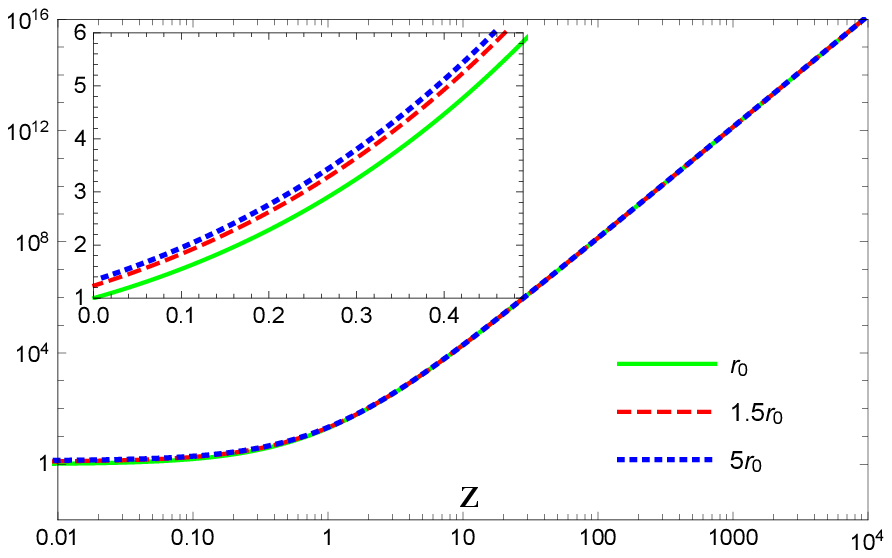}
} \subfigure[~$(\rho+p)/3H_0^2$ vs $z$] {\label{fig3c}\includegraphics[width=.32\textwidth]{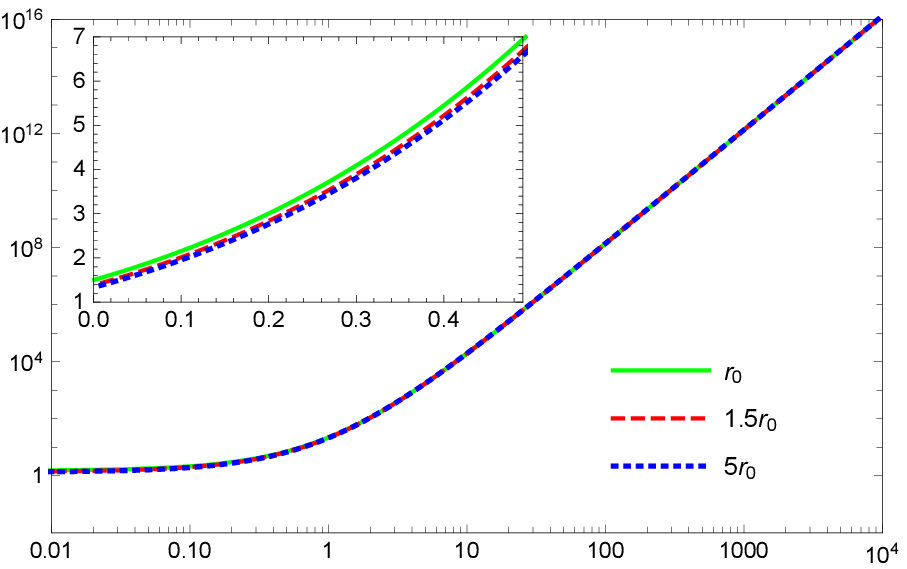} }
\caption{The behaviors of $\protect\rho $, $\protect\rho -\protect\tau $ and 
$\protect\rho +p$, respectively, versus $z$ for different values of $r$ in
the specific case of $\protect\omega_b=1/3$ with $U(\protect\phi)=%
\protect\phi^{2}$. Note that both the horizontal and vertical
axes are logarithmic.}
\label{fig3}
\end{figure*}

\begin{figure*}[htb!]
\centering\subfigure[~$\rho/3H_0^2$ vs $z$] {\label{fig4a}%
\includegraphics[width=.32\textwidth]{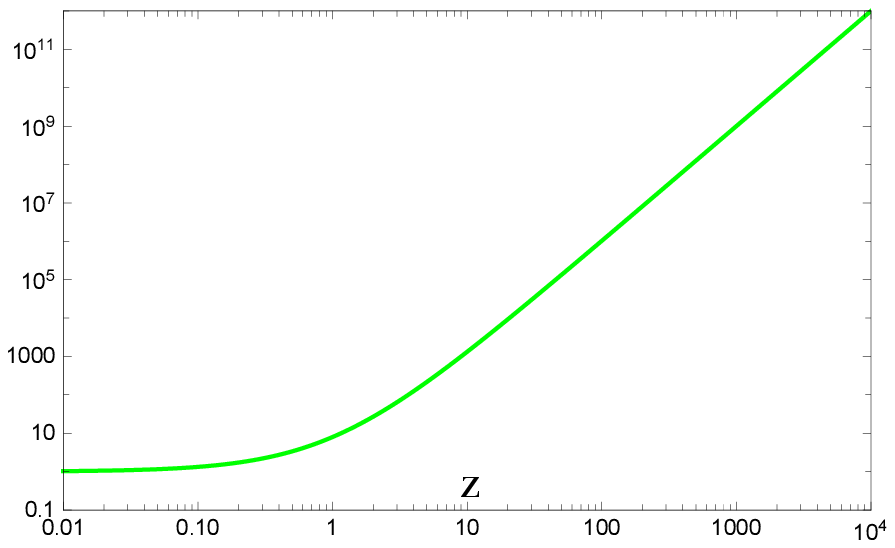} } 
\subfigure[~$(\rho-\tau)/3H_0^2$ vs $z$] {\label{fig4b}\includegraphics[width=.32\textwidth]{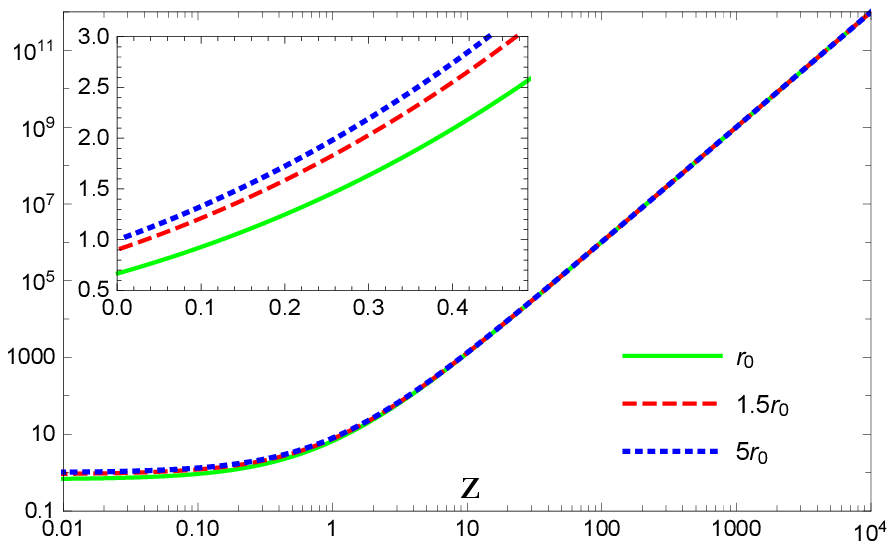}
} \subfigure[~$(\rho+p)/3H_0^2$ vs $z$] {\label{fig4c}\includegraphics[width=.32\textwidth]{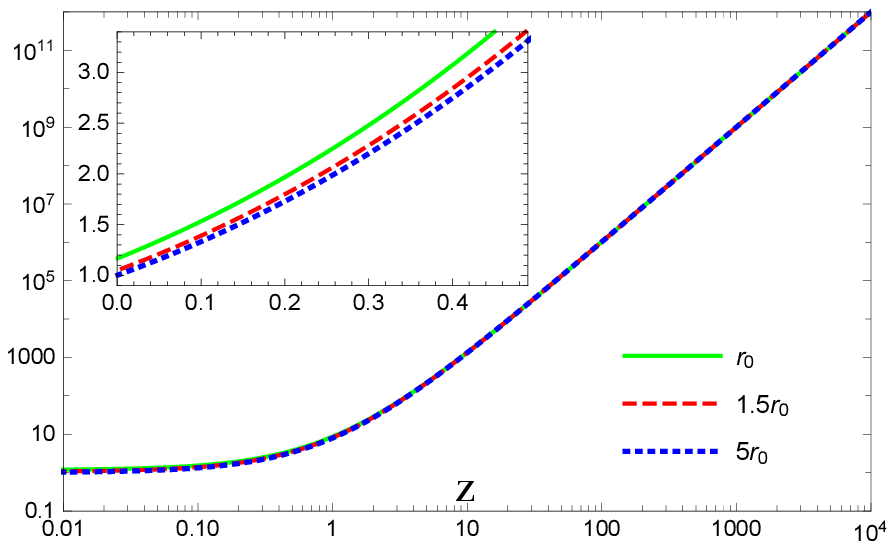} }
\caption{The behaviors of $\protect\rho $, $\protect\rho -\protect\tau $ and 
$\protect\rho +p$, respectively, versus $z$ for different values of $r$ in
the specific case of $\protect\omega_b=0$ with $U(\protect\phi)=%
\protect\phi^{2}$. Both horizontal and vertical axes are
logarithmic.}
\label{fig4}
\end{figure*}

\begin{figure*}[htb!]
\centering
\subfigure[~$\phi$ vs $z$] {\label{fig5a}\includegraphics[width=.32\textwidth]{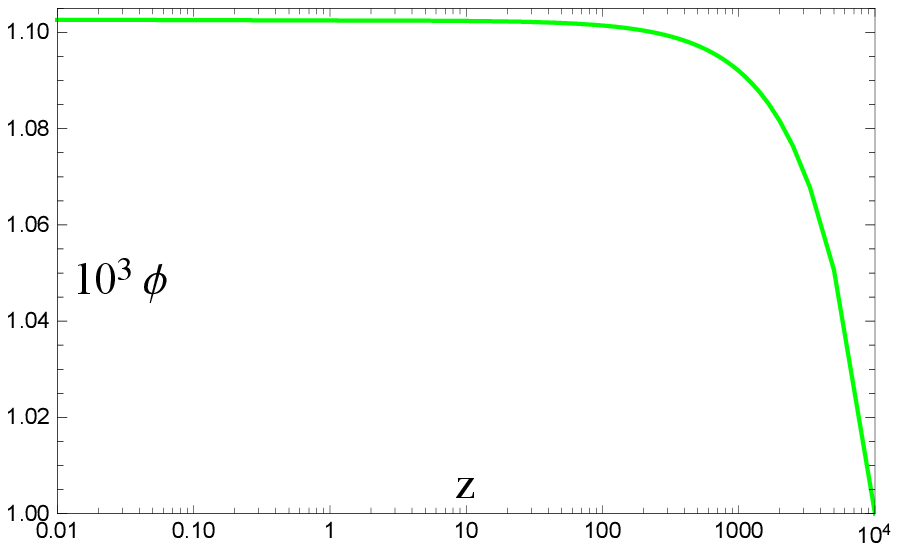}}
\subfigure[~$\phi'(a)$ vs $z$] {\label{fig5b}\includegraphics[width=.32\textwidth]{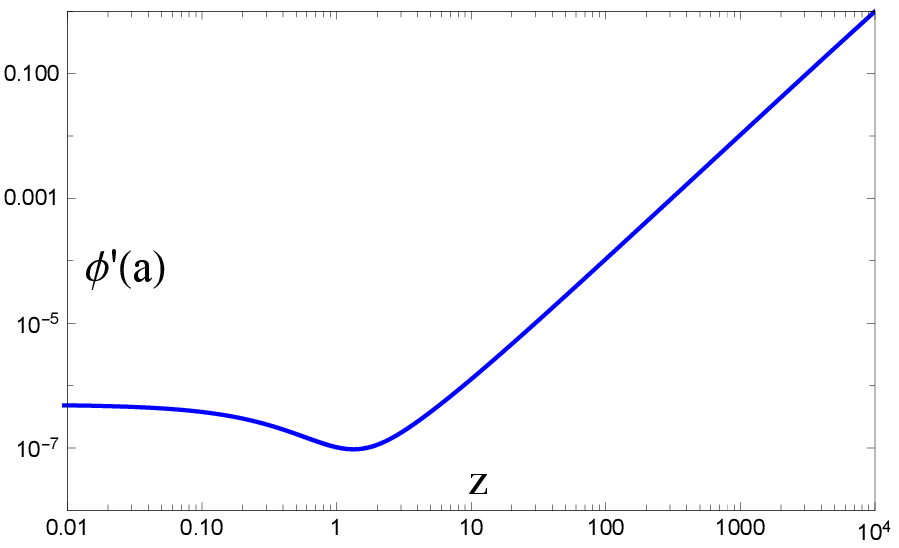}}
\subfigure[~$E$ vs $z$] {\label{fig5c}\includegraphics[width=.32\textwidth]{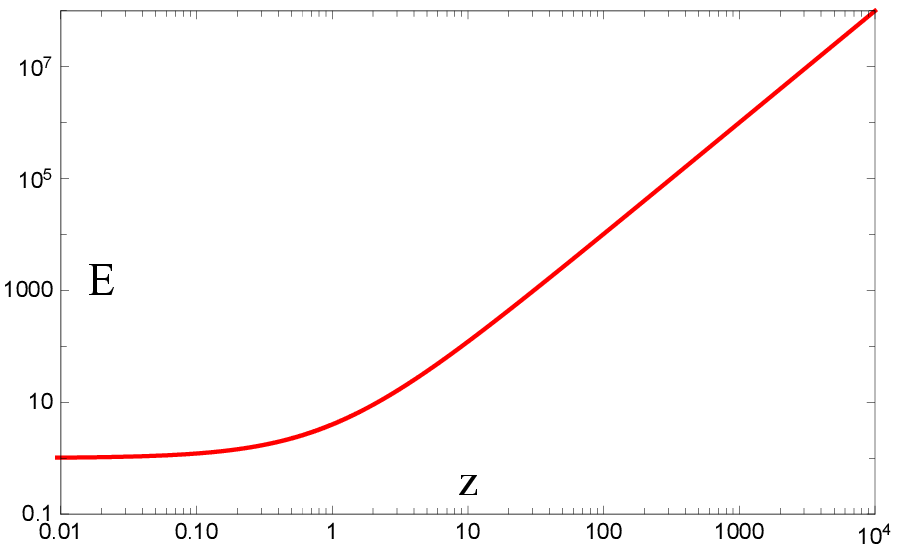}}
\caption{The behaviors of $\phi$, $\phi'(a)$ and $E$ vs $z$ for the traceless EMT case with $U(\protect\phi)=\protect\phi^{2}$. Note that both horizontal and vertical axes are logarithmic, except for the vertical axis of Fig. \ref{fig5a}.
See the text for more details.}
\label{fig5}
\end{figure*}

\begin{figure*}[htb!]
\centering\subfigure[~$\rho/3H_0^2$ vs $z$] {\label{fig6a}\includegraphics[width=.32\textwidth]{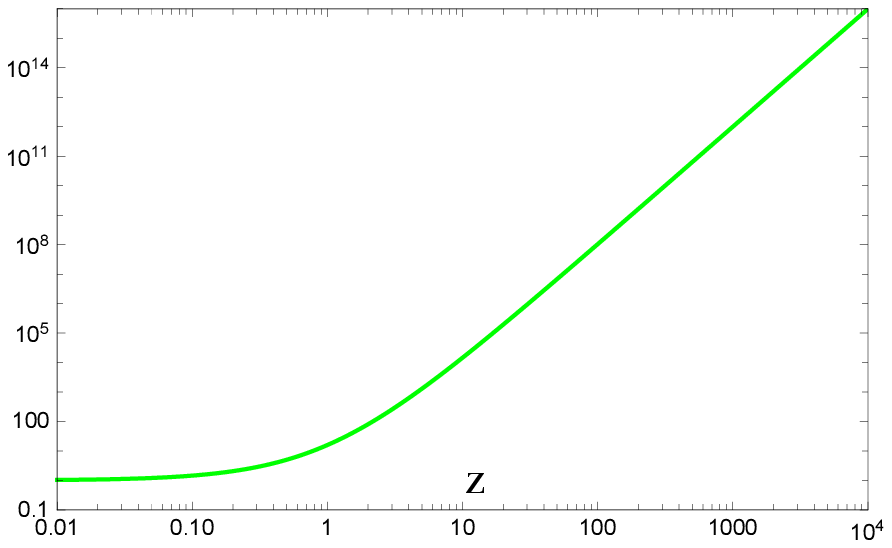} } 
\subfigure[~$(\rho-\tau)/3H_0^2$ vs $z$] {\label{fig6b}\includegraphics[width=.32\textwidth]{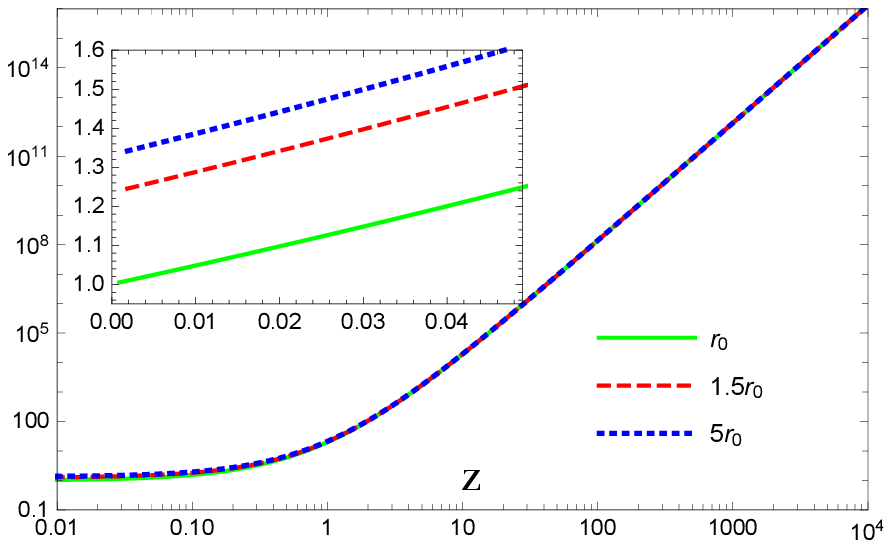}
} \subfigure[~$(\rho+p)/3H_0^2$ vs $z$] {\label{fig6c}\includegraphics[width=.32\textwidth]{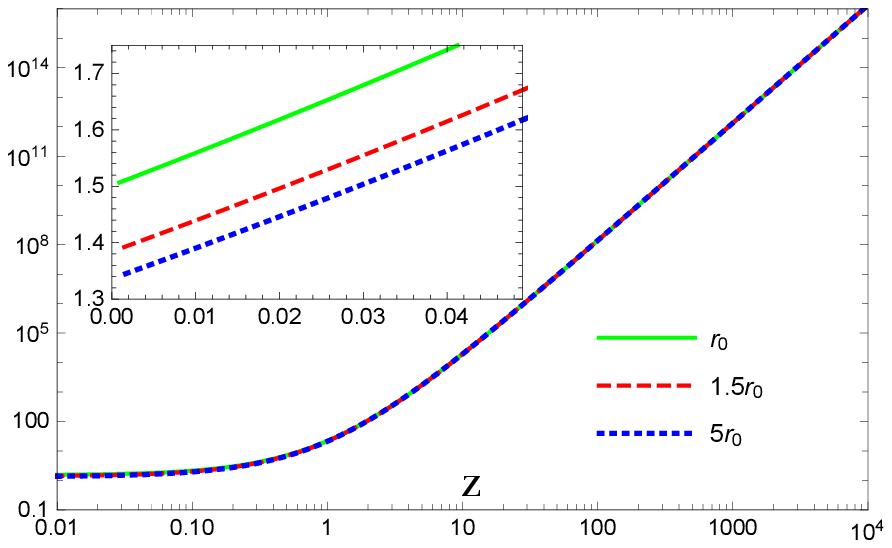} }
\caption{The behaviors of $\protect\rho $, $\protect\rho -\protect\tau $ and 
$\protect\rho +p$, respectively, versus $z$ for different values of $r$ for a wormhole with traceless EMT with $U(\protect\phi)=\protect\phi^{2}$. Both horizontal and vertical axes are logarithmic.}
\label{fig6}
\end{figure*}

In this section, we analyse evolving traversable wormhole geometries for the specific parameters $\omega _{b}=(-1, 1/3,0)$ considered above, as well as evolving wormholes with a traceless EMT, and study the NEC and WEC for the solutions obtained. 

Before going on further, we would like to comment on the potential for the scalar field $U(\phi)$ and its effects on the results. As Eqs. (\ref{rhotau}) and (\ref{rhop}) (and (\ref{rhoeq})-(\ref{pteq})) show, the quantities $\rho-\tau$ and $\rho+p$ (and consequently the NEC) are independent of the scalar field potential. However, it affects the behavior of $\rho$. We will consider a power law potential for scalar field $U(\phi)=\phi^\alpha$. There are some reasonable choice for $\alpha$, such as $\alpha=2$ and $\alpha=4$, inspired by a mass term and the Higgs potential, respectively. In fact, we emphasize that all the results below hold qualitatively for potentials of the form $\phi^2$ and $\phi^4$. For the sake of economy, we will present the results for the quadratic case, $U(\phi)=\phi^2$. In addition to this, to verify that other choices for $\alpha$ could also present reliable results, we consider the specific choice $\alpha=0.5$ in one of the cases.

\subsection{Specific case: $\omega _{b}=-1$}

Here, we consider the specific case $\omega _{b}=-\tau _{b}/\rho _{b}=-1$ with $U\left( \phi \right) =V\left(\phi \right) /3H_{0}^{2}=\phi^{0.5}$. Figure \ref{fig1a} depicts the behaviors of $\phi$ and $\phi^{\prime}$ with respect to the redshift $z$ $(=1/a-1)$.
In Fig. \ref{fig2}, the behaviors of $\rho $, $\rho -\tau $ and $\rho +p$ versus $z$ for
different values of $r$ are shown.
As one can see, $\rho - \tau$ is negative throughout the entire evolution. We emphasize that we have imposed different possible initial values and obtained the same result. Note that one may justify this outcome in the following manner. For this case with $\omega _{b}=-1$, Eq. (\ref{rhotau}) reduces to
\begin{equation}
\frac{\rho -\tau }{3H_{0}^{2}} =
- \frac{\left( 1+\phi (a)\right)r_{0}}{3a^{2}H_{0}^{2}r^{3}}
+a^{2}\left( \frac{\phi ^{\prime 2}(a)}{2\phi (a)}-\frac{\phi ^{\prime
\prime }(a)}{3}\right).
\end{equation}
Since $a\leq 1$, the term with a coefficient $1/a^2$ is dominant. This term is therefore negative, rendering $\rho -\tau $ negative. Thus, for this specific case, both the NEC and the WEC are violated.

\subsection{Specific case: $\omega _{b}=1/3$}

Figure \ref{fig1b} illustrates the behavior of  $\phi$ and $\phi^{\prime}$ versus $z$ for the specific case $\omega _{b}=1/3$. For this case, $E=a^{-2}$, $a(t) \propto t^{1/2}$ and we consider $U\left( \phi \right)=\phi^{2}$. The behaviors of $\rho $, $\rho -\tau $ and $\rho +p$ with respect to the redshift $z$ for different values of $r$ are displayed in Fig. \ref{fig3}, which shows explicitly 
that these quantities decrease as time evolves. However, they remain positive at all times and consequently
the NEC and WEC are always satisfied. This occurs for the wormhole throat as
well as for other wormhole radii. It is interesting to note that at a specified
time/redshift, the quantity $\rho -\tau $ increases for increasing values of
the radius, and the minimum value corresponds to the throat, as depicted by
Fig. \ref{fig3b}. On the other hand, $\rho+p$ decreases for increasing
values of the radius, and has a maximum at the throat (Fig. \ref{fig3c}).}

\subsection{Specific case: $\omega _{b}=0$}

Here we intend to study the energy conditions for the case $\omega _{b}=0$, for which $E=a^{-3/2}$ and $a(t)\propto t^{2/3}$. One can verify the behaviors of the scalar field $\phi$ and its derivative with respect to $a$ as functions of the redshift $z$ in Fig. \ref{fig1c}, where we consider the power law potential corresponding to scalar field as $U\left( \phi \right) =\phi ^{2}$.
The energy conditions are depicted in Fig. \ref{fig4}, where one verifies that $\rho $, $\rho -\tau $ and $\rho +p$ are positive at all times and for different values of spatial coordinate $r$. These quantities are also decreasing (increasing) functions with respect to time (redshift). Thus, the NEC and WEC are always satisfied for all wormhole radii including the wormhole throat. The behaviors of $\rho -\tau $ and $\rho + p $ at a specified time/redshift are analogous to the previous case with $\omega _{b}=1/3$, namely, the quantity $\rho -\tau $ ($\rho + p $) increases (decreases) for increasing values of the radius (see Figs. \ref{fig4b} and \ref{fig4c}).

\subsection{Wormholes with traceless EMT}

In this subsection, we consider the traceless energy-momentum tensor, with $T=
-\rho -\tau +2p=0$, which taking into account the dimensionless quantities defined above, provides the following differential equation
\begin{eqnarray}  \label{tracelessEMTb}
 -3 a^2 E(a)^2 \left[\phi ^{\prime \prime }(a)-\frac{\phi ^{\prime}(a)^{2}}{2 \phi (a)}\right]
-3 a E(a) \big[ a E^{\prime }(a)
\notag \\
+4 E(a) \big] \phi^{\prime}(a) 
	-6 E(a) \left[ a  E^{\prime }(a)+2 E(a)\right] 
\notag \\	
\times \left[1+\phi (a) \right]	+6 U(\phi)=0. 
\end{eqnarray}
Here, we solve the system of coupled differential equations (\ref{phidleom}) and (\ref%
{tracelessEMTb}), for $\phi $ and $E$, numerically. The behaviors of $\phi$, $\phi^{\prime}$ and $E$ versus $z$, taking into account the choice $U\left( \phi \right) =\phi^{2}$, are shown in Fig. \ref{fig5}. Note that $E$ tends to unity at the present time ($z \rightarrow 0$), as expected (Fig. \ref{fig5c}).
In Fig. \ref{fig6}, the behaviors of $\rho $, $\rho -\tau $ and $\rho +p$, with respect to $z$ for
different values of $r$ are displayed.
Note that $\rho $, $\rho -\tau $ and $\rho +p$ are decreasing (increasing)
functions of time (redshift) and are positive for all radii, as time evolves.
Thus, the NEC and WEC are satisfied at all times.
Moreover, as shown in Figs. \ref{fig6b} and \ref{fig6c}, the qualitative behaviors of $\rho -\tau $ and $\rho + p$ with respect to $r$ is analogous to the previous two cases, namely, $\omega _{b}=1/3$ and $\omega _{b}=0$, where the quantity $\rho -\tau $ ($\rho+p$) has a minimum (maximum) value at throat.

\section{Discussion and Conclusion}\label{conclusion} 

In the present paper, we have studied the evolution of dynamic traversable wormhole geometries in a FLRW background in the context of hybrid metric-Palatini gravity. This theory, which recently attracted much attention, consists of a hybrid combination of metric and Palatini terms and is capable of avoiding several of the problematic issues associated to each of the metric or Palatini formalisms (for more details, we refer the reader to Refs. \cite{Capozziello:2015lza,Harko:2018ayt,Harko:2020ibn}). For the evolving wormholes, we presented the components of the energy-momentum tensor that supports these geometries in terms of the model's functions, namely, the scalar field, the scale factor and the shape function (we considered a zero redshift function, for simplicity). Furthermore, we found specific wormhole solutions by considering a barotropic equation of state for the background matter, i.e., $\tau _{b} = - \omega _{b} \rho _{b}$, and considered particular equation of state parameters.

More specifically, we showed that for the specific cases of $\omega _{b}=1/3$ and $\omega _{b}=0$, the entire wormhole matter satisfies the NEC and WEC for all times. The latter cases are similar to the wormhole geometries analysed in the presence of pole dark energy \cite{KordZangeneh:2020jio}, however, there the WEC is violated at late times, i.e., the energy density becomes negative. Thus, the present results outlined in this work strengthen the varying dark energy models and may suggest that hybrid metric-Palatini gravity is a rather more promising model to explore.
In addition to the barotropic equation of state, we also studied evolving wormhole geometries supported by the matter with a traceless EMT. For this specific geometry, we discovered that both the NEC and the WEC are satisfied at all times as well. These results are extremely promising as they build on previous work that consider that the energy conditions may be satisfied in specific flashes of time \cite{Kar:1994tz,Kar:1995ss}.

An interesting astrophysical observational aspect on how to detect these wormholes would be to analyse the physical properties and characteristics of matter forming thin accretion disks around the wormhole geometries analysed in this work, much in the spirit of the analysis carried out in Refs. \cite{Harko:2008vy,Harko:2009xf,Harko:2009gc,Harko:2017fra}. In fact, specific signatures could appear in the electromagnetic spectrum, thus leading to the possibility of distinguishing these wormhole geometries by using astrophysical observations of the emission spectra from accretion disks. This would be an interesting avenue of research to explore.

\begin{acknowledgements}
We thank the referees for the constructive comments that helped us to significantly improve the paper.
MKZ would like to thank Shahid Chamran University of Ahvaz, Iran for
supporting this work. FSNL acknowledges support from the Funda\c{c}\~{a}o
para a Ci\^{e}ncia e a Tecnologia (FCT) Scientific Employment Stimulus
contract with reference CEECIND/04057/2017, and thanks funding from
the research grants No. PTDC/FIS-OUT/29048/2017, No. CERN/FIS-PAR/0037/2019 and 
No. UID/FIS/04434/2020.
\end{acknowledgements}


\end{document}